\documentclass[twoside,fleqn]{article}
\usepackage{amssymb,graphicx,espcrc2}

\newcommand{\txl}{T$\chi$L}

\newcommand{\xips}{\xi_{\rm ps}}

\newcommand{\mps}{m_{\rm ps}}

\newcommand{\mv}{m_{\rm v}}

\title{Exploiting finite-size-effects to simulate full QCD with light
  quarks---a progress report}

\author{B.~Orth\thanks{orth@theorie.physik.uni-wuppertal.de},
    N.~Eicker, Th.~Lippert, K.~Schilling, W.~Schroers, Z.~Sroczynski\\[1ex] 
    Department of Physics, Gau{\ss}stra{\ss}e 20, D-42097 Wuppertal, Germany}

\begin{document}

\begin{abstract}
We present a report on the status of the GRAL project {\em(Going
Realistic And Light)}, which aims at simulating full QCD with two
dynamical Wilson quarks below the vector meson decay threshold,
$\mps/\mv < 0.5$, making use of finite-size-scaling techniques.
\end{abstract}

\maketitle

\setcounter{footnote}{0}

\section{INTRODUCTION}%
The computer simulation of full QCD in the regime of light quark
masses, {\it i.e.} below the vector meson decay threshold, $R \equiv
\mps/\mv < 0.5$, represents one of the most interesting and yet long
standing goals of lattice QCD. As the algorithms and computer
resources of today are still inadequate for directly simulating
realistic quark masses without running into severe {\em finite size
effects} (FSE), GRAL will make a virtue of necessity by performing
simulations on {\em sequences} of small and medium sized lattices;
using {\em finite size scaling} techniques we will attempt to
extrapolate the results to the infinite volume prior to the chiral and
continuum extrapolations. In order to make contact with SESAM and
T$\chi$L data we use standard Wilson fermions.

\section{FIXING PHYSICAL PARAMETERS}%
Seeking to simulate QCD with light quarks our first task is to
determine an {\em operating point} ($\beta^{\rm op}$, $\kappa^{\rm
op}$) in parameter space with $R^{\rm op} < R_d = 0.5$. In order to
render simulations at this point feasible we specify $N_s=16$ as the
maximal spatial extent of a series of lattices employed. With the
number of lattice points in temporal direction held fixed at $N_t=32$,
the computational cost for $N_s=16$ thus sets an upper bound for the
costs of subsequent simulations on smaller lattices.

To be specific, we require the simulation with $N_s^{\rm op}=16$ at
$(\beta^{\rm op},\kappa^{\rm op})$ to yield meson masses with $R^{\rm
op}=0.4$ and a FSE of $z^{\rm op} = 0.2$. This particular value for
the finite size parameter \mbox{$z\equiv 1/(a\mps N_s)$} has been
chosen in view of a recent SESAM/T$\chi$L quark mass
analysis~\cite{eicker}, where $z \approx 0.2$ was found to mark the
onset of significant FSE's in the spectrum.

In summary, the envisaged properties of our target point
on a $16^3 \times 32$-lattice are as follows:
\[
 \newcommand{\cc}[1]{\multicolumn{1}{c}{\rule[-2mm]{0mm}{6mm} #1}}
 \renewcommand{\tabcolsep}{1pc} 
 \renewcommand{\arraystretch}{1.2} 
 \begin{array}{ccccccc} \hline
  \cc{\frac{\mps}{\mv}} & \cc{z} & \cc{\xips} & \cc{a\mps} & \cc{a\mv} &
   \cc{a [{\rm fm}]} & \cc{L\ [{\rm fm}]} \\
   \hline
   0.4 & 0.2 & 3.2 & 0.31 & 0.78 & 0.116 & 1.86 \\
   \hline
 \end{array}
\]
The lattice spacing $a \equiv a_\rho^l$ has been estimated on the
basis of the SESAM and T$\chi$L data at $\beta = 5.6$ and
$5.5$\footnote{For key parameters of the data see table 1
of~\cite{eicker}.}. Obviously, the price for sticking to $N_s=16$ is a
fairly coarse lattice. However, according to the scaling formulae
in~\cite{scaling_formula}, we expect a total cost of approximately
$0.2 \;\mathrm{TFlop/s}\cdot\mathrm{h}$ to produce 100 statistically
independent gauge field configurations at $(\beta^{\rm op},\kappa^{\rm
op})$, which is feasible.

\section{SEARCHING THE TARGET POINT}%
On the search for the bare run parameters $(\beta^{\rm op},\kappa^{\rm
op})$ we let ourselves be guided by a heuristic consideration. From
the known lattice spacings $a=0.095\;\mathrm{fm}$ at $\beta=5.5$ and
$0.078\;\mathrm{fm}$ at $5.6$, we deduce an estimate for $\beta^{\rm
op}$ from the renormalization group $\beta$-function. Assuming that we
are in the scaling regime we get $\partial\beta/\partial\ln a \approx
-0.46$, which leads us to expect that the presumed lattice spacing of
$a=0.116\;\mathrm{fm}$ at the target point will be attained at
$\beta^{\rm op} \approx 5.4$.



\begin{figure}[htb]
  \includegraphics[width=0.8\columnwidth,clip=]{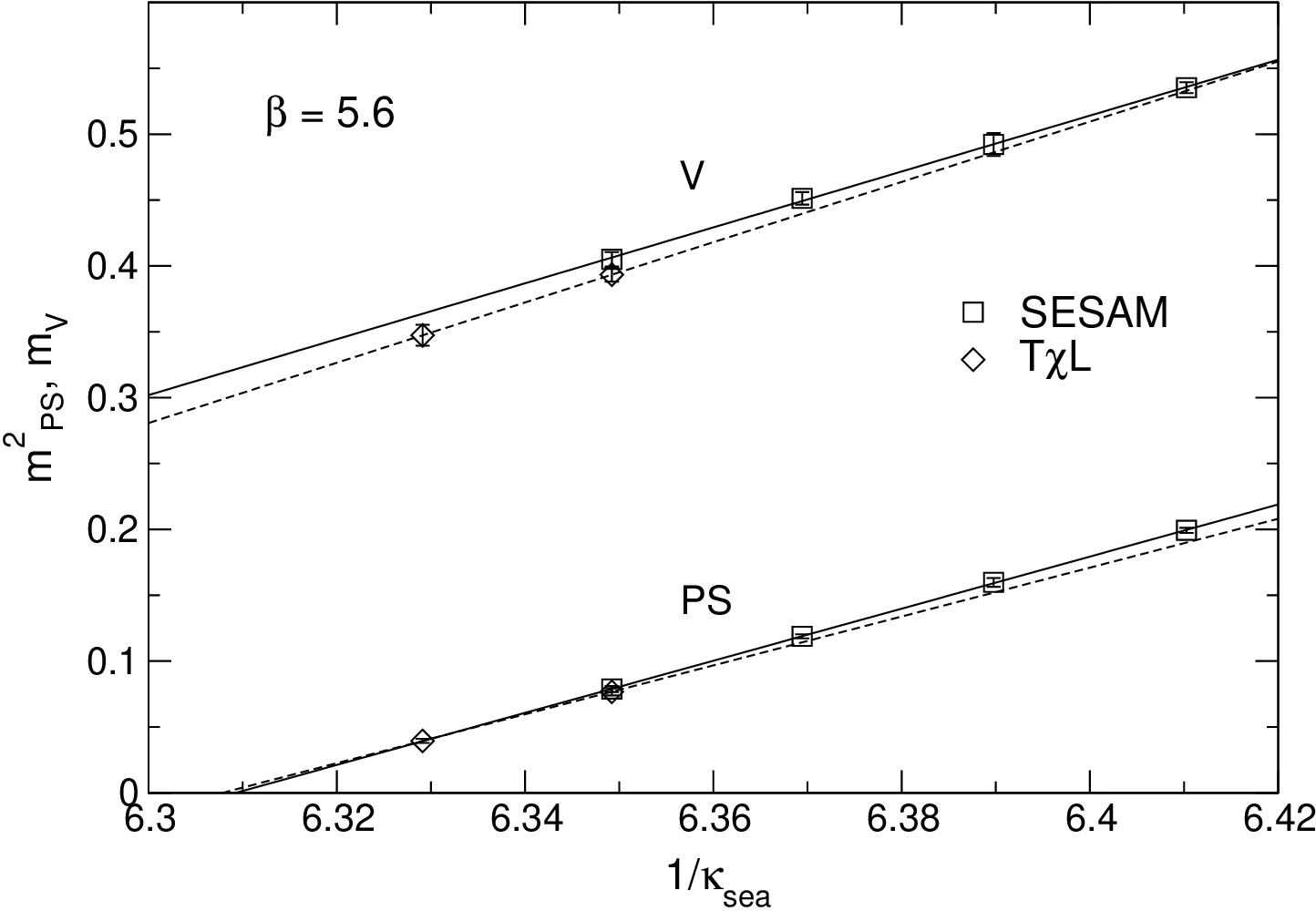}
  \vspace*{-.5cm}
  \caption{\label{fig:extrap_5.6} Chiral extrapolation of $\mv$ and $\mps^2$ for $\beta=5.6$.}
  \vspace*{-.5cm}
\end{figure}

\begin{figure}[htb]
 \includegraphics[width=.8\columnwidth,clip=]{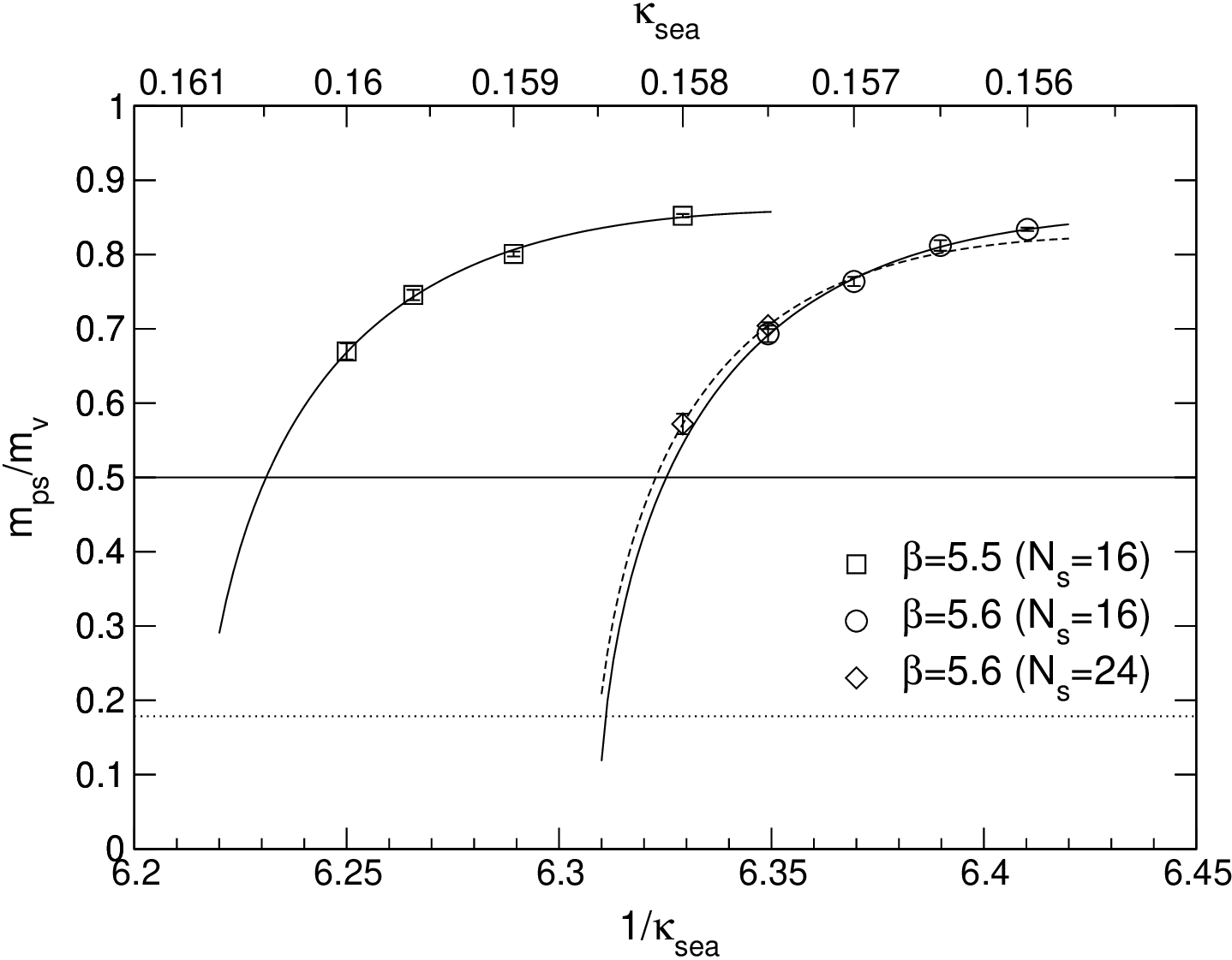}
 \includegraphics[width=.8\columnwidth,clip=]{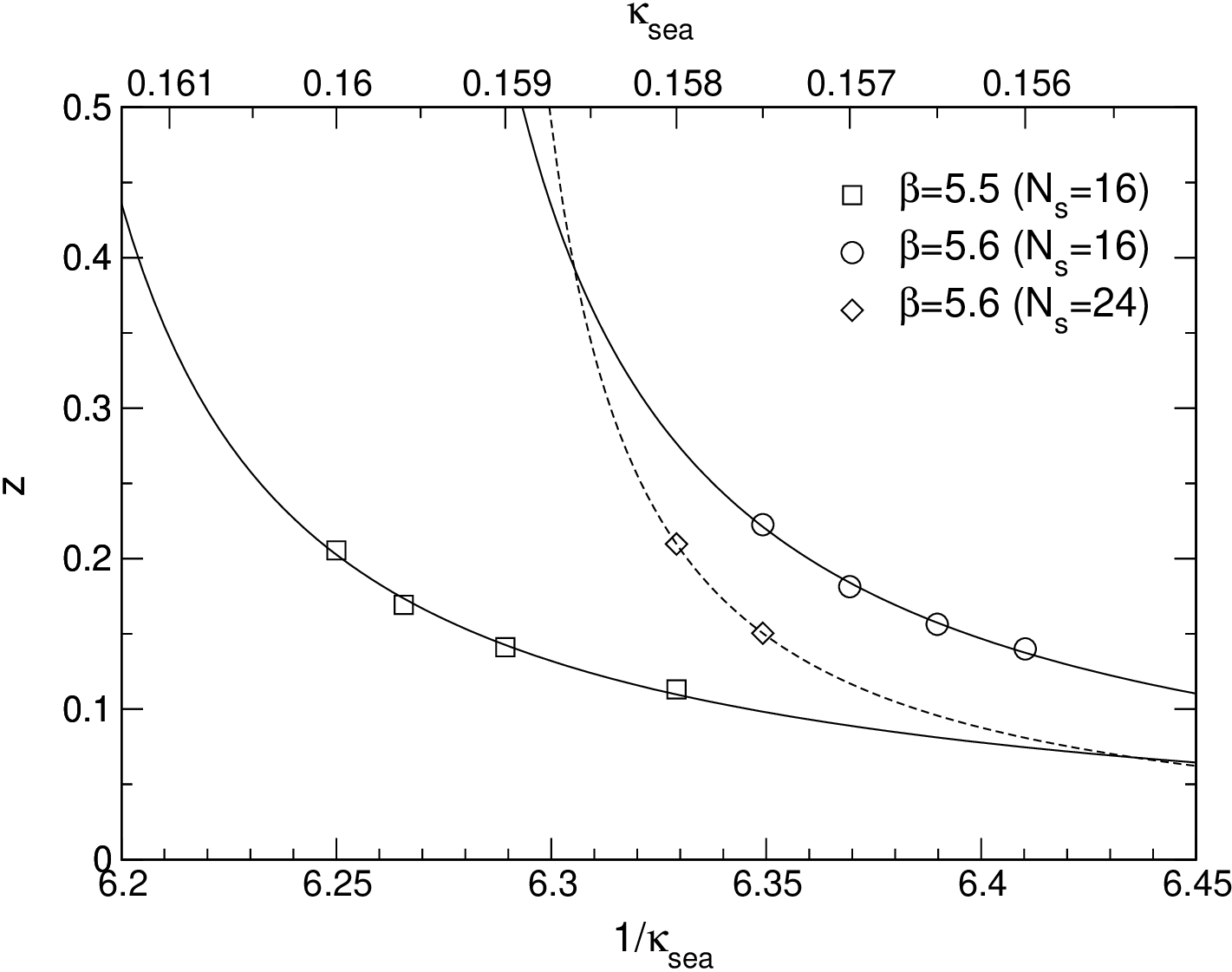}
 \vspace*{-.5cm}
 \caption{\label{fig:mpimrho_z} $R$ (a) and $z$ (b) as functions of $1/\kappa$.}
 \vspace*{-.5cm}
\end{figure}

With this in mind we turn to the SESAM/T$\chi$L spectral data. The
solid lines in fig.~\ref{fig:extrap_5.6} represent linear fits to
the $\mv$ and $\mps^2$ data at $\beta=5.6$ with $N_s=16$. The dashed
lines result from separate fits to the \txl\ data with $N_s=24$. We
use these fits (and corresponding ones for $\beta=5.5$) to establish
functional dependencies of $R$ and $z$ on $1/\kappa$, as shown in
fig.~\ref{fig:mpimrho_z}.

\begin{figure*}[!t]
  \includegraphics[width=\textwidth,clip=]{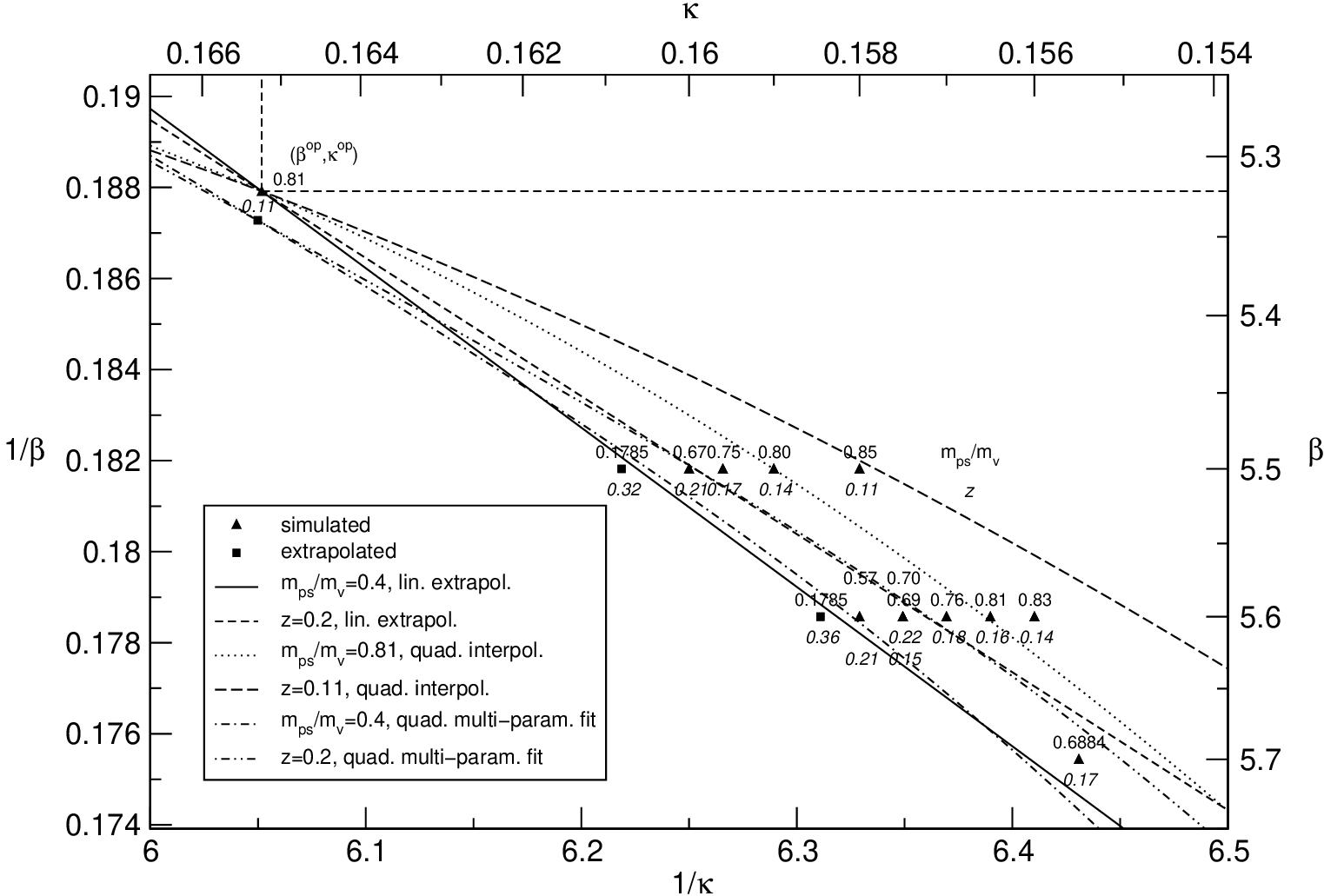}
  \vspace*{-1cm}
  \caption{\label{fig:beta_kappa} Lines of constant $\mps/\mv$ and $z$ in the $\beta$-$\kappa$ plane.}
\end{figure*}

We are now in the position to guesstimate the operating point by direct
extrapolation in the $\beta$-$\kappa$ plane. Setting out from the
functions of fig.~\ref{fig:mpimrho_z} we use a linear and a quadratic
ansatz to construct curves of constant $R$ and $z$ in that plane,
displayed in fig.~\ref{fig:beta_kappa}. The intersection of the
straight lines for $R \equiv 0.4$ (solid) and $z \equiv 0.2$ (dashed)
in the $\beta$-$\kappa$ plane delivers with $(\beta^{\rm op},
\kappa^{\rm op}) = (5.32, 0.1652)$ a first estimate for the operating
point.  Although $\beta^{\rm op}$ comes out close to the estimate from
the $\beta$-function, this result is not to be taken too seriously, as
preliminary results of a simulation performed with these action
parameters give $R(\beta^{\rm op},\kappa^{\rm op})=0.8$ and
$z(\beta^{\rm op},\kappa^{\rm op})=0.1$, which is off the expectation
by a factor of two in both cases. In order to illustrate qualitatively
the behavior of $R(\beta,\kappa)$, the long-dashed curve shows a
quadratic interpolation between $(\beta^{\rm op},\kappa^{\rm op})$
(having $R=0.8$) and the points with the same $R$ at $\beta=5.5$ and
$5.6$. (The dotted curve shows the same for $z(\beta,\kappa)$.)

A two-dimensional quadratic fit of all available (simulated and
extrapolated) $R(\beta,\kappa)$ and $z(\beta,\kappa)$, respectively,
to functions of the form $f(\beta,\kappa) = c_0 + c_1 \beta + c_2
\kappa + c_3 \beta^2 + c_4 \kappa^2 + c_5 \beta \kappa$ leads to the
coordinates $(\beta^{\rm op}, \kappa^{\rm op}) = (5.34, 0.1653)$.  The
dot-dashed curves in fig.~\ref{fig:beta_kappa} show the resulting
$\beta(\kappa)$ for $R\equiv 0.4$ and $z\equiv 0.2$.  As this second
guess for $(\beta^{\rm op},\kappa^{\rm op})$ deviates only little
from the first one we are still running at $\beta=5.32$,
but with increased values of $\kappa$. Production is carried out on
APEmille at DESY Zeuthen.

\section{SIMULATION PLAN}

For the smaller lattices ($N_s < 16$) we will use the cluster computer
ALiCE at the University of Wuppertal. To carry out a chiral
extrapolation, we will generate additional ensembles at equal $\beta$,
but larger $R$.  The step to a larger $\beta$ can be done either by
going to $N_s=18,20,22$ at $R=0.4$ and $z=0.2$, or by moving on to
$z>0.2$, depending on the results at the present simulation point.
We are aware of the fact that our operating point may be
located far in the strong coupling regime. On the other hand we might
possibly take advantage of the increased ``$\beta$-shifts'' at light
sea quark masses. Eventually we want to reach values of $\beta\ge
5.6$.

\begin{figure}[!b]
 \includegraphics[angle=-90,width=0.72\columnwidth,clip=]{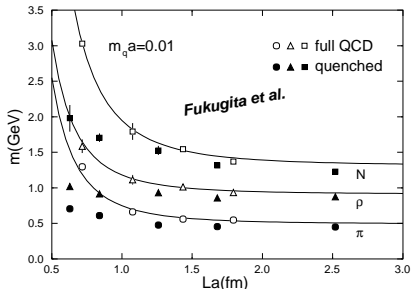}
 \vspace*{-.5cm}
 \caption{\label{fig:fukugita} Extrapolation to infinite volume.}
\end{figure}

For the extrapolation to the infinite volume limit we expect a formula
of the form $m = m_\infty + cL^{-\nu}$ to be
applicable~\cite{Fukugita:1992jj}. Fig.~\ref{fig:fukugita} shows
results of Fukugita {\it et al.}; the solid lines are fits with
$\nu=3$ to their full QCD data.

\section*{ACKNOWLEDGEMENTS}%
We thank I.~Montvay for important discussions. N.E.~is supported under
DFG grant Li701/3-1. B.O.~and W.S.~have been supported by the DFG
Graduiertenkolleg ``Feld\-theo\-re\-ti\-sche und nu\-me\-ri\-sche
Me\-tho\-den\dots''. Z.S.~is supported by the European Community's
Human potential program under HPRN-CT-2000-00145 Hadrons/Lattice QCD.

\bibliographystyle{h-elsevier}

\end{document}